\documentclass[reprint, amsmath,amssymb, aps]{revtex4-2}
\usepackage{graphicx}
\usepackage{xcolor}
\usepackage{float}
\usepackage{dcolumn}
\usepackage{bm}
\usepackage{natbib}
\usepackage{hyperref}
\usepackage{braket}
\usepackage{amssymb}
\usepackage{comment}
\usepackage{hyperref}
\hypersetup{
    colorlinks=true,
    linkcolor=blue,
    filecolor=magenta,      
    urlcolor=cyan,
    pdftitle={Overleaf Example},
    pdfpagemode=FullScreen,
    }
\usepackage{tikz}
\usepackage{circuitikz}
\usepackage{tikz,pgfplots}
\usetikzlibrary{shapes.geometric}
\usetikzlibrary{arrows.meta,arrows,shapes,positioning,calc}
\begin{document}

\preprint{APS/123-QED}

\title{Test of  Transitivity in Quantum Field theory using Rindler spacetime}
\author{Sashideep Gutti}
\email{sashideep@hyderabad.bits-pilani.ac.in}
\author{Akhil U Nair}
\email{p20200473@hyderabad.bits-pilani.ac.in}
\author{Prasant Samantray}
\email{prasant.samantray@hyderabad.bits-pilani.ac.in}

\affiliation{Department of Physics, Birla Institute of Technology and Sciences-Pilani \\Hyderabad, 500078, India}

\date{\today} 

\begin{abstract}
In a quantum field theoretic setting, we consider a system $\cal{M}$ and it's subsystem $R_1$. Let $R_2$ be a subsystem of $R_1$. We therefore have $R_2\subset R_1 \subset \cal{M}$ with the symbol $\subset$ implying a quantum subsystem. The reduced state in $R_2$ due to a quantum state in $\cal{M}$ can be found in two ways, a) by tracing out the unobserved degrees of freedom in $\cal{M}$ and b) by first finding the reduced state in $R_1$ and then tracing over unobserved degrees of freedom in $R_1$. In this letter we address the question whether both the methods yield the same reduced state in $R_2$. To this end we consider for $\cal{M} $  Minkowski spacetime with a massless scalar field in its vacuum state. For $R_1$ we consider a standard  Rindler wedge. For $R_2$ we consider a Rindler wedge shifted  to the right of $R_1$ by a distance $\Delta$. We find the reduced state in $R_2$ using two independent ways: i) evaluating of the reduced state from the vacuum state in $\cal{M}$, yielding a thermal density matrix, ii) by first evaluating the reduced state in $R_1$ from $\cal{M} $ yielding  a thermal state in $R_1$, and subsequently evaluate the reduced state in $R_2$ in that order of sequence. To that end, we devise a method which involves cleaving the Rindler wedge $R_1$ into two domains such that they form a thermofield double. One of the domains aligns itself along the wedge $R_2$ while the other is a diamond shaped construction between the boundaries of $R_1$ and $R_2$. We show that both these independent methods yield two different answers for the particle content in $R_2$. We discuss the possible implications of our result, with special focus on the quantum states outside a non-extremal black hole formed by collapsing matter. 
\end{abstract}

\maketitle

\begin{figure}[htb]
		\centering
		\begin{tikzpicture}[scale=.4]			
			\draw[<->,line width=1] (-10,0) -- (10,0)
			node[below] at(9,-0.5){$X_M$};
			\draw node[black] at (-5.9,5.2) {M};
			\draw[<->,line width=1] (-7,-5) -- (-7,5) node[ ] at (-8.9,5.5) {$T_M$};
			\draw[line width=.5,black](-2.5,-4.5)--(-7,0)--(-2.5,4.5) node[black] at (-3,5.2) {\scriptsize $R_1$};
			\draw[line width=.5,black](0.5,-4.5)--(-4,0)--(0.5,4.5)node[black] at (0.3,5.2) {\scriptsize $R_2/RR$};
			\draw[line width=.5,black](3.5,-4.5)--(-1,0)--(3.5,4.5)node[black] at (3.9,5.2) {\scriptsize $R_3/RRR$};
			\draw[line width=.5,black](6.5,-4.5)--(2,0)--(6.5,4.5)node[black] at (6.9,5.2) {\scriptsize $R_4$};
			\draw[line width=.5,black](9.5,-4.5)--(5,0)--(9.5,4.5)node[black] at (9.5,5.2) {\scriptsize $R_5$};
			\draw[line width=1.5,blue](-5.5, -1.5)--(-4, 0)--(-5.5, 1.5)--(-7,0)--(-5.5,-1.5);
			\draw[line width=1.5,blue](-2.5,-1.5)--(-4, 0)--(-2.5,1.5)--(-1,0)--(-2.5,-1.5);
			\draw[line width=1.5,blue](0.5,-1.5)--(-1,0)--(0.5,1.5)--(2,0)--(0.5,-1.5);
			\draw[line width=1.5,blue](3.5,-1.5)--(2,0)--(3.5,1.5)--(5,0)--(3.5,-1.5);
			\draw node[black] at (5,-6) {$R_5 \subset R_4 \subset R_3 \subset R_2 \subset R_1 \subset M$};
			
		\end{tikzpicture}
		\caption{Diamond Necklace: Minkowski spacetime with Rindler wedge-$R_1$, $R_2$, $R_3$. The Diamond regions are $RR-L$, $RRR-L$....}
		\label{diamondnecklace}
	\end{figure}
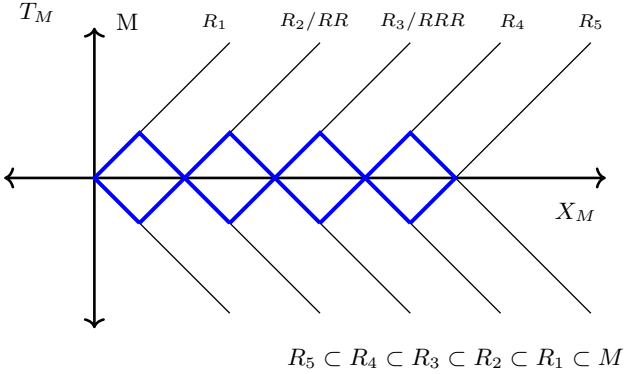

\emph{ Introduction:} Recent astrophysical observations  have propelled the study of black holes to the forefront of cutting edge research, both from a theoretical standpoint as well as in the context of observational cosmology. However, there remain quite a number of results regarding black hole physics that still await experimental confirmation - Hawking radiation, information retrieval  are yet to receive experimental validation which might come sooner than expected. Additionally from a theoretical perspective, extensive work is being carried out in the fields of black 
hole entropy, entanglement, quantum information  as reviewed in \cite{maldacena,witten1,bousso,suvrat,carlip}.\\ 
Unsurprisingly, Rindler spacetime, which happens to be the near horizon limit of non-extremal black holes has been extensively used to understand various properties of the event horizon. Entanglement entropy, Hawking radiation, and horizon properties can be investigated in basic models of this space time before being treated in more realistic scenarios like black hole spacetimes \cite{higuchi}.\\
It is well established that if we define a quantum field theory (say a massless scalar field) in Minkowski space and let the quantum state of the field be in vacuum, then a uniformly accelerated observer perceives the same vacuum to be filled with particles that are thermally populated. The Rindler spacetime is viewed as a quantum subsystem of the Minkowski spacetime and therefore the reduced state of Minkowski vacuum in Rindler spacetime is thermal. \\
In this letter, we point out an important caveat that is missed out in many calculations involving quantum fields in a subsystem or quantum field theory in curved background., that can lead to results that are not generic and are valid as special cases. In general terms the problem we address is this:  In a quantum field theoretic setting we consider a system M. A quantum  subsystem of M is $R_1$ and a quantum subsystem of $R_1$ is $R_2$. $R_2$ is therfore a subsystem of $M$ also.  To determine the reduced state in a subsystem from the quantum state of the full system, we trace over the unobserved degrees of freedom.  We can arrive at the reduced state in $R_2$ via two independent methods: 1) we interpret $R_2$ as a subsystem of M and trace over unobserved degrees of freedom of $R_2$ in M. 2) we interpret $R_2$ as a subsystem of $R_1$ which is in turn a subsystem of M. So $R_2$ is a sub-subsystem of M. We find reduced state of $R_2$ from the reduced state of R1. We therefore trace out degrees of freedom outside $R_1$ first and then trace out degrees of freedom outside $R_2$ but belongs to $R_1$. 
We justify in the letter that both the methods need not yield the same answer implying that specific criteria regarding entanglement between the subsystems might be needed for the answers to match.  The reduced state may indeed depend on the sequence of how the tracing is carried out. In the model we consider below we show explicitly that both the methods yield two different answers.  
\\
\\
{\it{Problem of Transitivity:}} Here we define the transitivity problem in quantum field theory within the framework of Rindler spacetime. We consider two-dimensional Minkowski space M with coordinates $T_M,X_M$. We consider a Rindler wedge ($R_1$) with standard interpretation in the figure (\ref{wedge}), with $\tau_1, \xi_1$ corresponding to Rindler coordinates. This wedge is such that the bifurcation point is at the origin (O in figure (\ref{wedge})) of the Minkowski spacetime.  A second Rindler wedge ($R_2$) is also considered, where the bifurcation point (O$_1$) is moved to the right by $\Delta$, as shown in the figure (\ref{wedge}. Let the quantum state of a scalar field in Minkowski space be in vacuum $\ket{0}_M$. The wedge $R_1$, with the scalar field constitute a quantum subsystem of Minkowski space, and the reduced quantum state in $R_1$ has a thermal spectrum with a density matrix $\rho_{R_1 M}$. As for the wedge $R_2$, we observe that it is a Rindler wedge shifted to the right by a $\Delta$. Once again the quantum  state in $R_2$ is a thermal state with density matrix $\rho_{R_2M}$. Here the subscript $2$ indicates the wedge ($R_2$) state, and the second subscript $M$ indicates the initial state, representing the Minkowski state. As $R_2$ is a quantum subsystem of $R_1$, the reduced quantum state of $R_2$ can be estimated from $\rho_{R_1 M}$ and is denoted as $\rho_{R_2 R_1}$. We take the thermal density matrix in $R_1$ given by $\rho_{R_1}$ and  estimate the reduced density matrix in $R_2$. The question is whether the quantum state deduced this way be the same as that which is obtained as a reduced state from Minkowski vacuum, i.e is $\rho_{R_2M}=\rho_{R_2R_1}$? In other words, is the quantum state transitive? The reduced state $\rho_{R_2M}$ involves tracing over the degrees of freedom on the $T_M=0$ line from $X_M=-\infty $ to $X_M=0$ whereas, the reduced state $\rho_{R_2R_1}$ is obtained in two steps - first tracing over the degrees of freedom on $T_M=0$ line  from $X_M=-\infty$ to $X_M=0$ to obtain $\rho_{R_1}$ from Minkowski vacuum and then tracing from $X_M=0$ to $X_M=\Delta$ (as shown in the figure (\ref{wedge})) to obtain $\rho_{R_2R_1}$. This letter aims to check whether the reduced quantum state is independent of the tracing details. This question is therefore of fundamental importance in understanding the behavior of quantum fields and quantum information content (when one estimates the Von Neuman entropy of the quantum state in $R_2$ as $S_{R2M}=-tr(\rho_{R_2M}ln(\rho_{R_2M}))$ as compared to $S_{R2R1}=-tr(\rho_{R_2R_1} ln(\rho_{R_2R_1}))$). This also has implications for understanding final quantum state in exterior of a black hole. This is elaborated in the last section. \\
\\
{\it{Quantum Mechanical situation:}} We first present a few remarks in the quantum mechanical scenario for completeness. We consider a quantum mechanical system consisting of $N$ coupled oscillators in their ground state along the lines of analysis done in \cite{srednicki},\cite{bombelli}. If we denote the coordinates of the oscillators by $(x^1,x^2...x^N)$ and consider the reduced density matrix of the system obtained by tracing out first 'M'  out of the N coupled oscillators, then the final reduced state will depend only on $(x^{M+1}, x^{M+2}...x^N)$. It was explicitly shown in \cite{srednicki} that we obtain a mixed state. If we now ask the question whether the mixed state depends on the sequence of the tracing over the coordinates $(x^1...x^M)$, i.e.  if we first trace out K oscillators and then L oscillators such that $K+L=M$ for various positive values of K and L. Does the reduced state depends on the order in which the tracing is performed? It is easy to verify using simple calculation that the reduced state of $N-M$ quantum system is independent of the order in which the tracing  operation is done. The density matrix obtained by tracing out $(x^1...x^M)$ is therefore invariant. However, the situation in quantum field theory is more subtle. \\
\\
{\it{The set-up:}}  Consider a massless scalar field in Minkowski spacetime $\cal{M}$, which is in its vacuum state. Based on the standard Bogoliubov method, it is trivial to compute the reduced state in $R_2$ from the vacuum state in $\cal{M}$. The challenging part is estimating the reduced state in $R_2$ from the thermal state in $R_1$.\\
\emph{Prelude:} The preprint \cite{kinjalkpaddy} discusses the question about how a vacuum state in $R_1$ appears in $R_2$. Below, we summarize the relevant coordinates for formulating our question. We use coordinates $(T_M, R_M)$ for two dimensional Minkowski spacetime $\cal{M}$; $(\tau_1,\xi_1)$ for the coordinates in the wedge $R_1$; and $(\tau_2,\xi_2)$ for the coordinates in the wedge $R_2$. Listed below are the relations between the various coordinates, $T_M=e^{a\xi_1}\sinh(a\tau_1)/a=e^{a\xi_2}\sinh(a\tau_2)/a$ and $X_M=e^{a\xi_1}\cosh(a\tau_1)/a=e^{a\xi_2}\cosh(a\tau_2)/a+\Delta$ . Where `$a$' indicates the acceleration parameter in each of the wedges $R_1$ and $R_2$ and $\Delta$ indicates the shift of the wedge $R_2$ from $R_1$ along the common $X_M-axis$ as shown in the figure (\ref{wedge}). The metric in Minkowski spacetime is $ds^2=-dT_M^2+dR_M^2$. The metric in $R_1$ is given by, $ds^2=e^{2a\xi_1}(-d\tau_1^2+d\xi_1^2)$. Similarly for $R_2$ the metric is given by  $ ds^2=e^{2a\xi_2}(-d\tau_2^2+d\xi_2^2)$. Using the above, one can easily deduce the horizon structure (causal boundaries) of $R_1$ and $R_2$ as illustrated in the diagram \ref{wedge}. Separate acceleration parameters $a_1$ and $a_2$ for the first and second wedges can be specified, but this does not change the qualitative aspects of the results. So for simplicity, we assume both acceleration parameters to be the same. \\
\begin{figure}[htb]
		\centering
		\begin{tikzpicture}[scale=.6]			
			\draw[<->,line width=2] (-1,0) -- (10,0)
			node[below] at(9.5,-0.3) {$X_M$-axis} node[] at (2.5,0.5){$\Delta$};
			\draw node[black] at(2.5,1) {$RR-L$};
			\draw node[black] at (1.7,3.6) {Minkowski};
			\node[black] at(-0.4,-0.5) {\large $O$};
			\node[black] at(5,-0.7) {\large $O_1$};
			\draw[>=triangle 45, <->] (0.3,0.1) -- (4.7,0.1);
			\draw[<->,line width=2] (0,-5) -- (0,5) node[ ] at (-1.5,5) {$T_M$-axis};
			\draw[line width=1.5,red](4,-4)--(0,0)--(4,4)node[above,black]{$R_1$};
			\draw[line width=1.5,purple](2.5,-2.5)--(0,0)--(2.5,2.5);
			\draw[line width=1.5,purple](2.5,-2.5)--(5,0)--(2.5,2.5);
			\draw[line width=1.5,blue](9,-4)--(5,0)--(9,4);
			\draw node[black] at (9.5,2.6) {$R_2,  RR-R$};
			\draw node[black] at (7,-4.8) {$R_2 \subset R_1 \subset M$};
		\end{tikzpicture}
		\caption{Minkowski spacetime (M) with Rindler wedge-$R_1$ and Rindler wedge-$R_2$. $R_2$ and $RR-R$ coincide.}
		\label{wedge}
	\end{figure}
Null coordinates  provide us with greater insight into our problem. Here the null rays are defined as $U_M=T_M-X_M$ and $V_M=T_M+X_M$. Similarly, we define $u_i=\tau_i-\xi_i$ and $v_i=\tau_i+\xi_i$, where $i=1,2$ for the wedges $R_1$ and $R_2$ respectively. It is easy to deduce the relationship between them, as shown below,
\begin{equation}
U_M=-\frac{e^{-au_1}}{a}=-\frac{e^{-au_2}}{a}-\Delta,
\label{urelation}
\end{equation}
\begin{equation}
V_M=\frac{e^{av_1}}{a}=\frac{e^{av_2}}{a}+\Delta.
\label{vrelation}
\end{equation}
The coordinate range for all the null rays are $-\infty<U_i, V_i<\infty$ (where $i$ takes values $M, 1$ and $2$). This range makes it easy to see that the horizon for $R_1$ is given by ($u_1=\infty$, $v_1=-\infty$) which in Minkowski null coordinates is given by ($U_M=0$,$V_M=0$). Similarly for the wedge $R_2$, the horizons can be found to be given by ($u_2=\infty$, $v_2=-\infty$) which in Minkowski null coordinates is given by ($U_M=-\Delta$,$V_M=\Delta$) and in terms of $R_1$ null coordinates, the horizons of $R_2$ map to ($u_1=-\ln(a\Delta)/a$, $v_1=\ln(a\Delta)/a$). Hereafter we choose the value of $\Delta=1/a$ so that the horizons of the wedge $R_2$ pass through the origin of $R_1$, $(\tau_1=0,\xi_1=0)$.  We evaluate the relationship between $(u_1,v_1)$ and $(u_2,v_2)$ in the near horizon limit for the wedge $R_2$. In the limit of $u_2 \rightarrow -\infty$, in which case, from equation \ref{urelation}, we get $u_1=u_2$.  In the limit $v_2\rightarrow -\infty$, from \ref{vrelation} we get $e^{av_1}=e^{av_2}+1$. Taking logarithm on both sides and power expanding $\ln(1+x)$ for small x, we get $v_1=e^{av_2}/a$ (these limits also apply to early time behavior of $(u_2, v_2)$). Similarly, $v_1=v_2$ and $u_1=-e^{-au_2}/a$ represents the late time behavior. The above relations between null rays of $R_1$ and $R_2$ can be summarized as follows: $u_1=-e^{-au_2}/a$, $v_1=e^{av_2}/a$. This observation is reminiscent of the relation between Kruskal coordinates and Schwarzschild light cone coordinates or between Minkowski and the Rindler coordinates. Later, we will demonstrate that this near-horizon relation plays a crucial role in particle content.\\ 
\\
{\it{Cleaving of Rindler chart $R_1$. The rich substructure of Rindler spacetime:}} We now take the following detour that lets us estimate the reduced state based on the relation between the null coordinates near the horizon of $R_2$. The strategy is to cleave the Rindler spacetime $R_1$ into two parts such that they form a thermofield double of each other. This split regions play the same role in the context of Rindler spacetime $R_1$ as Rindler Left and Right wedges play in the context of Minkowski spacetime.
In \cite{kolekar2}, the authors have defined an interesting spacetime  called Rindler-Rindler (RR) spacetime. We use these coordinate to carry out the cleaving of $R_1$ spacetime. The RR coordinates $(\tau_{rrr},\xi_{rrr})$ are defined as $\tau_1=e^{a\xi_{rrr}}\sinh(a\tau_{rrr})/a$ and $\xi_1=e^{a\xi_{rrr}}\cosh(a\tau_{rrr})/a$. We have used the subscript `$rrr$' for Rindler-Rindler-Right since we show later how we can define Rindler-Rindler-Left coordinates. In \cite{kolekar2}, the authors have constructed quantum field theory in Rindler-Rindler-Right ($RR-R$) spacetime and showed that the vacuum state in $R_1$ appears as a thermally populated state in $RR-R$ spacetime. The metric in these coordinates is conformal and is given by $ds^2=e^{2 a (\xi_1 +\xi_{rrr})}(-d\tau_{rrr}^2+d\xi_{rrr}^2)$.  We show below that the $RR-R$ spacetime and  $R_2$ wedge share the same horizons and therefore are two different coordinate systems for the same region of spacetime. In the figure (\ref{wedge}), we indicate that $R_2$ and $RR-R$ occupy the same wedge region. To demonstrate this,  we define null coordinates of $RR-R$ spacetime and find the map between null coordinates of $R_1$ and $RR-R$. From the definition of coordinates, we easily derive that $u_1=-\frac{e^{-au_{rrr}}}{a}$ and $v_1=\frac{e^{av_{rrr}}}{a}$ where $(u_{rrr}=\tau_{rrr}-\xi_{rrr})$ and $(v_{rrr}=\tau_{rrr}+\xi_{rrr})$. Considering the range of null rays $-\infty<u_{rrr},v_{rrr}<\infty$,  we arrive at the fact that the relevant  boundary/horizon of $RR-R$ spacetime is given by ($u_{rrr}=\infty$, $v_{rrr}=-\infty$). This corresponds to $(u_1=0,v_1=0)$, showing that the $RR-R$ spacetime and $R_2$ spacetime share the same boundary. In fact this relation between Rindler wedges and Rindler-Rindler spacetimes can be made more general by choosing different acceleration parameters 'a' for the wedges given a shift $\Delta$. This implies that given a shifted wedge, the cleaving of the $R_1$ spacetime can be done by choosing an appropriate acceleration parameter 'a' such that $RR-R$ coincides with the shifted wedge. \\
 Our goal is to estimate the features of particle content in $R_2$ due to the thermal state in $R_1$ which is in turn  a reduced state from pure vacuum state of the scalar field in Minkowski spacetime.  We observe that in the near horizon (of $R_2$) behavior of the two coordinate systems $(u_2,v_2)$ and $(u_{rrr},v_{rrr})$ are equal due to the fact that $u_1=-\frac{e^{-a u_2}}{a}=-\frac{e^{-a u_{rrr}}}{a}$, $v_1=\frac{e^{av_2}}{a}=\frac{e^{av_{rrr}}}{a}$. Close to the horizon, the  coordinates of $R_2$  and $RR-R$ converge. We can also arrive at the same conclusion from the expression for the metric in $RR-R$ spacetime. We can write the metric in terms of null coordinates as $ds^2=e^{2 a (\xi_1 +\xi_{rrr})}(-d\tau_{rrr}^2+d\xi_{rrr}^2)=e^{ a (v_1-u_1 +v_{rrr}-u_{rrr})}(-du_{rrr}dv_{rrr})$. From these expressions, we can see that near the horizon of $R_2$ ($v_1->0,u_1->0)$, the two coordinate systems $(u_2,v_2)$ and $(u_{rrr},v_{rrr})$ coincide. We make an assumption that the particle content in the reduced state is crucially dependent on the near horizon behavior of the modes. We therefore estimate the particle content of $R_2$ in the Rindler coordinates by considering the isometry between both the coordinates in the near horizon limit and the causal structure of $RR-R$ region. 
We note that $\partial_{\tau_{rrr}}$ is not a killing vector and the metric in $RR-R$ coordinates is not stationary. Nevertheless the vector $\partial_{\tau_{rrr}}$, in the near horizon limit as well as late/early times, aligns itself along the Killing vector $\partial_{\tau_2}=X_M\partial_{T_M}+T_M\partial_{X_M}-\Delta \partial_{T_M}$ due to the coordinates$(\tau_2,\xi_2)$ coinciding with ($\tau_{rrr}, \xi_{rrr})$. We now construct the Rindler-Rindler-Left wedge. 
\\
{\it{Causal Diamond construction for Rindler-Rindler-Left Chart}} : We define the 'left wedge' $RR-L$ (Rindler-Rindler-Left) using the definition, $(\tau_{rrl},\xi_{rrl})$ such that $\tau_1=-e^{a\xi_{rrl}}\sinh(a\tau_{rrl})/a$ and $\xi_1=-e^{a\xi_{rrl}}\cosh(a\tau_{rrl})/a$. The null coordinates with proper range determine the boundary of this left wedge. This region is like a diamond in the Minkowski spacetime diagram as shown in the figure (\ref{wedge}). Surprisingly this region is the thermofield twin of the $RR-R$ region as justified subsequently in this article.  As suggested in \cite{kolekar2} one can think of a series of Rindler-Rindler-Rindler-Rindler.... frames. Now by constructing $RRR-L, RRRR-L...$ spacetimes we can get these diamonds and arrive at a diamond necklace structure in Minkowski spacetime as seen in the figure (\ref{diamondnecklace}). In fact by choosing different acceleration parameters `$a$', one can vary the size of the diamonds in the necklace without the physicist worrying about the budget! As of now, reduced quantum states and entropy aspects of these diamond regions have not been studied.\\
	\\
{\it{Modes}}: For the Rindler wedge $R_1$, the standard modes defined on the positive Minkowski coordinate $X>0$ form a complete set and this lets one to define the quantum field in the wedge $R_1$. Below we show that the modes defined in $RR-R$ and $RR-L$ have the exact same relation with the wedge $R_1$, as that of the relation shared by right and left versions of the Rindler wedges with Minkowski spacetime. We start by defining the modes in $RR-R$ spacetime. The massless Klein-Gordon equation for the $RR$ spacetime (with the understanding that $(\tau_{RR}, \xi_{RR})=(\tau_{RR-R},\xi_{RR-R})$ in right wedge and $(\tau_{RR}, \xi_{RR})=(\tau_{RR-L},\xi_{RR-L})$ for the left wedge) is, $-\partial_{\tau_{RR}}^2{\Phi}+\partial_{\xi_{RR}}^2{\Phi}=0$. The positive frequency modes (defined with respect to $\partial_{\tau_{rr}}$) with support on $RR-R$ wedge can be found to be, $f_k^{RR-R}=\left(e^{-i\omega_k\tau_{rr}+ik\xi_{rr}}\right)/\left(4\pi\right)$ with $\omega_k=|k|$. Similarly the positive frequency modes (defined with respect to $-\partial_{\tau_{rr}}$) with support in $RR-L$ are found to be $f_k^{RR-L}=\left(e^{i\omega_k\tau_{rr}+ik\xi_{rr}}\right)/\left(4\pi\right)$. These modes, together with their complex conjugates form a complete set in $RR-R$ and $RR-L$ spacetimes respectively. We note that the Cauchy surface for $R_1$ is $\tau_1=0$ (positive $X_M$ axis), while $\tau_{rrr}=0$ is the Cauchy surface for $RR-R$ ($X_M=1/a$ to $X_M=\infty$) and similarly, $\tau_{rrl}=0$ for $RR-L$ spacetime (from $X_M=0$ to $X_M=1/a$). The scalar product for the modes has the standard definition and is given by,
\begin{equation}
    \langle f,g \rangle = -i \int_{\Sigma}{}\left(f\partial_{\mu}g^* - g^* \partial_{\mu}f\right)n^{\mu}\sqrt{h}d^3 x.
\end{equation}
where $\sqrt{h}$ is the square root of the determinant of the induced metric on the spacelike hypersurface ($e^{a\xi_1}$ for $\tau_1=0$ in $R_1$ and $e^{a(\xi_1+\xi_2)}$ for the hypersurface $\tau_{RR-R}=0$ and $\tau_{RR-L}=0$ for $RR-R$, $RR-L$ spacetime) and $n^{\mu}$ is the unit normal to the hypersurface  (in all the cases discussed in this article, $n^{\mu}$ turns out to be non zero only for the time component and has magnitude $1/ \sqrt{h}$). The normalization of the modes is defined with respect to the above norm. The creation and annihilation operators for the modes $f_k^{RR-R}$ and $f_k^{ RR-L}$ are $\left(b_k^{\dagger RR-R}, b_k^{RR-R}\right)$ and $\left(b_k^{\dagger RR-L}, b_k^{RR-L}\right)$ respectively. The vacuum in $RR-R$ and $RR-L$ is defined as  $\left(b_k^{RR-R}\ket{0}_{RR-R}=0\right)$ and $\left(b_k^{RR-L}\ket{0}_{RR-L}=0\right)$ respectively. In the paper \cite{kolekar1}, it is shown that if we take a vacuum state for the Rindler spacetime $R_1$, the state $\ket{0}_{R1}$ yields a mixed thermal state in $RR-R$ with the temperature $a/2\pi$. This calculation is done by finding out the Bogoliubov coefficients between the modes of $R_1$ and $RR-R$. We rederive the same fact by defining equivalent Unruh-modes for the $RR$ spacetime below. By using these Unruh modes we can determine the particle content of $RR-R$ given a thermal state in $R_1$.\\ 
\\
{\it{Unruh modes:}} The discussion follows the standard treatment as given in \cite{birrell} and \cite{carroll}. It is easily shown that $f_k^{RR-R}=\left( \left(-u_1\right)^{i\omega_k/a}a^{i\omega_k/a}\right)/\left(\sqrt{4\pi \omega_k}\right)$ in the $RR-R$ region. In the $RR-L$ region we can show that  $f_{-k}^{*RR-L}=\left(e^{\pi\omega_k/a}\left(-u_1\right)^{i\omega_k/a}a^{i\omega_k/a}\right)/\left(\sqrt{4\pi \omega_k}\right)$  which implies that the modes can be analytically continued into each other over the entire $\tau_1=0$ plane. Just as in the case of Rindler spacetime, we can now construct the Unruh modes that are well defined in both the left and right Rindler-Rindler wedges. The normalized Unruh modes take the form similar to the Minkowski Unruh modes,
\begin{equation}
    h_k^1=\frac{e^{\pi\omega_k/2a}f^{RR-R}_k+e^{-\pi\omega_k/2a}f^{\dagger RR-L}_{-k}}{\sqrt{2\sinh(\pi\omega_k/a)}},
    \label{unruhmode1}
\end{equation}
\begin{equation}
    h_k^2=\frac{e^{\pi\omega_k/2a}f^{RR-L}_k+e^{-\pi\omega_k/2a}f^{\dagger RR-R}_{-k}}{\sqrt{2\sinh(\pi\omega_k/a)}}.
    \label{unruhmode2}
\end{equation}

The quantum field in the region $R_1$ can therefore be expressed in terms of both the modes as,
\begin{equation}
    \phi=\int dk (d_k^1h_k^1+d_k^{\dagger1}h_k^{* 1}+d_k^2h_k^2+d_k^{\dagger2}h_k^{*2}).
\end{equation}
where $d_k^1, d_k^2$, $\left(d_k^{\dagger 1}, d_k^{2\dagger}\right)$ are the annihilation (creation) operators corresponding to the two Unruh-modes. Both the annihilation operators operate on the vacuum state of $R_1$ to yield $d_k^1\ket{0}_{R_1}=0$, $d_k^2\ket{0}_{R_1}=0$. To rederive the result in \cite{kolekar2} using these Unruh modes, we express the Rindler-Rindler operators as,
\begin{equation}
    b_k^{RR-R}=\frac{e^{\pi\omega_k/2a}d^{1}_k+e^{-\pi\omega_k/2a}d^{\dagger 2}_{-k}}{\sqrt{2\sinh(\pi\omega_k/a)}},
    \label{bk}
\end{equation}
\begin{equation}
    b_k^{RR-L}=\frac{e^{\pi\omega_k/2a}d^{2}_k+e^{-\pi\omega_k/2a}d^{\dagger 1}_{-k}}{\sqrt{2\sinh(\pi\omega_k/a)}}.
    \label{bkdagger}
\end{equation}
 To find out how Rindler vacuum ($R_1$) appears in $RR-R$, we evaluate the expression for the expectation of the number operator on $R_1$ vacuum state and obtain,
\begin{equation}
_{R_1}\bra{0}b_k^{\dagger RR-R}b_k^{RR-R}\ket{0}_{R_1}=\frac{1}{e^{2\pi\omega_k/a}-1},
\end{equation}
 showing the Planckian distribution with temperature given by $a/2\pi$. The same can be shown for $RR-L$ case.
  \\
  \\
{\it{Particle content in near horizon limit of $R_2$: }} With this background, We now attempt to address the central issue raised in this article. We begin by expressing Rindler vacuum in terms of its thermofield double,
\begin{equation}
    \ket{0_{R_1}}=\Pi_k A^2_k\sum_ne^{-n\pi \omega_k/a}\ket{n}_{RR-L}\otimes \ket{n}_{RR-R},
\end{equation}
with $A_k^2$ being the normalization constant. We have started with Minkowski vacuum $\ket{0}_M$. The reduced state in Rindler wedge $R_1$ is a mixed thermal state with temperature $a/(2\pi)$.  In the papers \cite{kolekar1}, \cite{kolekar3}  the reduced state in Rindler spacetime ($R_1$) is estimated when the Minkowski space is in a  thermal state with temperature $T'$. The reduced state in $R_1$ is shown to be non thermal and they derive a nice explicit analytical expression for the particle number density in $R_1$. Our analysis closely follows \cite{kolekar1}, \cite{kolekar3} owing to the conformal nature of the metric as well as scalar field considered here being massless.  We consider the analysis for one set of Unruh-Rindler particles with modes $h^1_k$. The density matrix for a thermal state in $R_1$ is given by  
\begin{equation}
    \rho_{R_1}=\Pi_k B_k^2\sum_m \frac{e^{-2\pi \omega_k m/a}}{m!}(d_k^{1\dagger})^m\ket{0}_{R_1}\bra{0}_{R1} (d^1_k)^m ,
    \label{rhoth}
\end{equation}
with $B_k^2$ normalization constant.
Following the notation in \cite{kolekar1},\cite{kolekar2},  and using  equations (\ref{bk}), (\ref{bkdagger}) we express $d_k^1$ in terms of RR-R and RR-L creation and annihilation operators as,
\begin{equation}
    d^{\dagger1}_k=\frac{b^{\dagger RR-R}_k-\bar{\mathcal{Q}}b^{RR-L}_{(-k)}}{\bar{\mathcal{P}}(1-\bar{\mathcal{Q}})} ,
    \label{dk}
\end{equation}
with $\bar{\mathcal{Q}}=e^{-\pi\omega_k/a}$ and $\bar{\mathcal{P}}=(e^{\pi\omega_k/2a})/\sqrt{2 sinh(\pi\omega_k/2a)}$. Using the expression for $(d^{\dagger 1}_k)^m$ from Appendix in \cite{kolekar2}, obtained by binomially expanding equation (\ref{dk}), and tracing over the resultant density matrix over the RR-L states, we obtain the reduced density matrix given by (after equating both the temperatures in \cite{kolekar2} to $a/2\pi$),
\begin{equation}
\begin{split}
    \rho_{RR-R}=\Pi_kC_k^2\sum_p\left(2-e^{-{2\pi\omega_k/a}}\right)^p\\
   \times e^{-2\pi p\omega_k/a} \ket{p}_{RR-R}\bra{p}_{RR-R} .
\end{split}
\end{equation}
Here $C_k$ is the normalization constant which is calculated to be  $C_k=(1-e^{-\omega_k 2\pi/a})$ by imposing the condition that the trace of density matrix equals one. We are interested in the number density in $RR-R$ spacetime and this can be evaluated from the expression
\begin{equation}
\begin{split}
    \langle N \rangle_k=tr\left[\rho_{RR-R} b^{\dagger RR-R}_kb^{RR-R}_k\right]\\ =\frac{1}{e^{2\pi\omega_k/a}-1}\left(2+\frac{1}{e^{2\pi\omega_k/a}-1}\right) .
    \end{split}
\end{equation}
This is therefore the particle content in the shifted Rindler spacetime $R_2$ in the near horizon limit. The above particle spectrum is obviously not Planckian, and is therefore non thermal. If transitivity were to hold in quantum field theory in the presence of horizons, we would have had a thermal spectrum with temperature given by $a/2\pi$. But instead our particle estimate yields a non thermal spectrum and therefore the result is indicative of the fact that transitivity is lost in the problem. The conclusion being that the reduced quantum state in a quantum field theoretic setting can depend on the order in which the tracing out is carried out. \\
\\
\textit{Consequences and Discussion:}  In situations where transitivity is absent, the reduced state of a subsystem is specific to one particular sequence of tracing out procedure and does not imply a general answer. Failure of transitivity raises many questions. Some of the pertinent questions in the model discussed in the letter are as follows. If we refine the wedges between $R_1$ and $R_2$ with $N$ number of wedges, say $W_1,W_2..W_N$  such that $R_2 \subset W_N \subset..W_i...\subset W_1 \subset R_1$, one may ask the question - What is the reduced state in $R_2$ due to intermediate N wedges  calculated from $R_1$? 

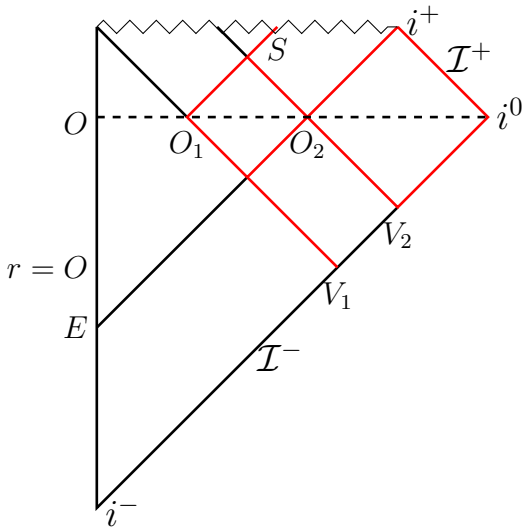
\begin{figure}[h]
    \centering
    \begin{tikzpicture} [scale=0.4]
    \draw[line width = 1, black] (-3,3)--(-3,-13)--(7,-3);
    \draw[line width = 1, black] (2,2)--(1,3);
    \draw[line width = 1, black] (-3,3)--(0,0);
    \draw[line width = 1, red] (3,3)--(2,2);
    \draw[line width = 1, red] (2,-2)--(5,-5);
    \draw[line width = 1, black] (-3,-7)--(2,-2);
    \draw[line width = 1, red] (2,2)--(0,0)--(2,-2)--(4,0)--(2,2);
    \draw[line width = 1, red] (7,3)--(4,0)--(7,-3)--(10,0)--(7,3);
    \draw[line width = 1, black,dashed] (-3,0)--(10,0);
    \draw[decorate,decoration=zigzag] (-3,3) -- (7,3);
    \node[black,left] at(-3,-0.2) {\large $O$};
    \node[black,below] at(3,2.9) {\large $S$};
    \node[black, left] at (-3,-7) {\large $E$};
    \node[black,below] at(0,-0.2) {\large $O_1$};
    \node[black,below] at(4,-0.2) {\large $O_2$};
    \node[black,below] at(3.1,-7.2) {\Large $\mathcal{I}^{-}$};
    \node[black,above] at(9.4,1.2) {\Large $\mathcal{I}^{+}$};
    \node[black,below] at(5,-5.2) {\large $V_1$};
    \node[black,below] at(7,-3.2) {\large $V_2$};
    \node[black,right] at(7,3.2) {\Large $i^+$};
    \node[black,right] at(10,0.1) {\Large $i^0$};
    \node[black,right] at(-3,-13.1) {\Large $i^-$};
    \node[black,left] at(-3,-5) {\large $r=O$};
    \end{tikzpicture}
    \caption{Collapse of two null shells at $V_1$ and $V_2$ with mass $M_1$ and $M_2$. The static exterior is $V_2$, $i^0$, $i^+$, $O_2$. The diamond region with diagonal $O_1$ and $O_2$ is also static with mass $M_1$ in it's metric written in Schwarzschild coordinates.}
    \label{collapse}
\end{figure}

 Can one estimate the form of the reduced state after $N$ iterations? Is there an asymptotic expression for the reduced state (in terms of particle content, density matrix )  as $N$ tends to infinity? Can one deduce an asymptotic formula if it exists?\\
How does this Rindler model discussed in the letter have relevance in the real world situation?
To that end, we claim that our model using shifted Rindler wedges becomes relevant  in the realistic scenario involving dynamical horizons. In \cite{ashtekar2002}, it is proved that in the collapse of generic matter,  the dynamical horizon is a spacelike hypersurface. This fact is crucial in connecting our shifted Rindler situation to the realistic gravitational collapsing scenario. In the figure \ref{collapse}, we describe a two step process. Two null shells with null coordinates given by $V_1$ and $V_2$ (with masses $M_1$ and $M_2$) collapse to form the black hole. The arbitrary shift between the two Rindler wedges in the Letter now gets naturally determined by the details of the two null shells as can be seen in the figure \ref{collapse}. The dynamical horizon is spacelike between $O_1$ and $O_2$ and is isolated between $O_2$ and $i^+$ ( The light ray $E-O_2-i^+$ is the event horizon). The Schwarzschild exterior with mass $(M_1+M_2)$ is given by the region enclosed by vertices $V_2,i^0,i^+,O_2$. The static exterior is a quantum subsystem of the wedge $S,O_1,V_1, i^0,i^+$ and is similar to the shifted wedge situation discussed in the article. The diamond shaped patch with diagonal vertices being $(O_1,O_2)$ is also a static patch with mass $M_1$. The relevant reduced state in the exterior region can be obtained from a "in" vacuum state by either tracing over the line $O,O_2$, or via tracing over two steps viz ($O,O_1$) and then $(O_1, O_2)$ similar to the shifted Rindler scenario discussed in the letter. Do both these methods yield the same state in the exterior? Is there an asymptotic form if we consider many intermediate wedges?  These questions in the context of framework developed in the article are left for future considerations. An information theoretic proof towards the problem discussed in the letter can offer insights into the subtle nature of entanglement between the systems and their subsystems that  obey transitivity. These questions are open and left for future considerations.
 \\
 
 \bibliography{Ref}

\end{document}